\documentclass[conference]{IEEEtran}
\IEEEoverridecommandlockouts
\usepackage{cite}
\usepackage{amsmath,amssymb,amsfonts}
\usepackage{algorithmic}
\usepackage{graphicx}
\usepackage{textcomp}
\usepackage{xcolor}
\usepackage{verbatim}
\usepackage{url}
\usepackage{pgfplots}          
\usepackage{pgfplotstable}
\usepackage{filecontents}
\usepackage{tikz}              
\usepackage{caption}           
\pgfplotsset{compat=newest}   
\usepgfplotslibrary{groupplots}
\usepgfplotslibrary{statistics}

\def\BibTeX{{\rm B\kern-.05em{\sc i\kern-.025em b}\kern-.08em
    T\kern-.1667em\lower.7ex\hbox{E}\kern-.125emX}}

\newcommand{\bra}[1]{\langle #1|}
\newcommand{\ket}[1]{|#1\rangle}

\newcommand{\braket}[2]{\left\langle #1\lvert#2\right\rangle}

\newcommand\defn[1]{\textsl{#1}}

\def\cH{\mathcal H}	
\def\C{\mathbb{H} }	

\begin{document}

\title{Adiabatic training of VQAs for entanglement detection\\

}

\author{
\IEEEauthorblockN{Ernesto Acosta}
\IEEEauthorblockA{\textit{Dpt. Computer Science and AI} \\
\textit{University of Granada}\\
Granada, Spain \\}
\and
\IEEEauthorblockN{Carlos Cano}
\IEEEauthorblockA{\textit{Dpt. Computer Science and AI} \\
\textit{University of Granada}\\
Granada, Spain \\
}
\and
\IEEEauthorblockN{Daniel Manzano}
\IEEEauthorblockA{
\textit{QTCG.}\\
\textit{Dpt. Electromagnetism}\\ \textit{and Condensed Matter} \\
\textit{University of Granada}\\
Granada, Spain \\
}
\and
\IEEEauthorblockN{Guillermo Botella}
\IEEEauthorblockA{\textit{Dpt. Computer Architecture}\\ \textit{and Automation} \\
\textit{Complutense University}\\
Madrid, Spain \\
}
}

\maketitle

\begin{abstract}
This paper introduces a hybrid quantum framework for training Variational Quantum Algorithms (VQAs) aimed at addressing the fundamental problem of entanglement detection using adiabatic quantum computing. This problem is of foundational significance in quantum information science and has been approached through both classical and quantum methodologies. We propose an adiabatic-inspired optimization strategy for training VQAs, which combines a discretization of the parameter space with a hierarchical segmentation scheme. The optimization problem is formulated as a Quadratic Unconstrained Binary Optimization (QUBO) model to leverage the capabilities of quantum annealing. Preliminary results are compared against classical neural networks and conventional VQA training techniques.  
The findings aim to stimulate further investigation and discussion within the scientific community regarding the efficacy and applicability of the proposed approach. 
\end{abstract}


\section{Introduction}

Entanglement stands as one of the most profound and counterintuitive features of quantum mechanics, fundamentally distinguishing it from classical theories. Originally brought to prominence by Einstein, Podolsky, and Rosen in their seminal critique of the completeness of quantum theory \cite{einstein:pr35}, and subsequently identified by Schrödinger as {\it the characteristic trait of quantum mechanics, the one that enforces its entire departure from classical lines of thought} \cite{schrodinger:mpcps35}, entanglement has evolved from a conceptual curiosity into a cornerstone of modern quantum science. Its significance extends far beyond foundational debates, playing a central role in the development of quantum technologies. However, despite its pivotal role, the theoretical and practical characterization of entanglement remains an ongoing challenge. In particular, the formulation of effective separability criteria and reliable entanglement measures continues to be a central focus of research \cite{horodecki:rmp09}. The task of detecting and quantifying entanglement in arbitrary quantum states—especially in high-dimensional or multipartite systems—remains largely unresolved, highlighting the need for novel approaches and deeper insights.

Variational Quantum Algorithms (VQAs) have arisen as a pivotal class of hybrid quantum-classical algorithms. These algorithms have been extensively studied not only for accomplishing tasks that are quantum in nature, but also for classical optimization and classification tasks~\cite{Cerezo2021}. VQAs employ a parameterized quantum circuit, or ansatz, governed by a set of training parameters that are optimized to minimize (or maximize) a cost function. Therefore, VQAs can be used for classification when they are trained to approximate a mapping from an input (quantum state) to its corresponding output (label), by means of measurements and circuit parameter adjustments.

While VQAs have become increasingly popular, recent research has focused on developing training algorithms beyond standard gradient descent in order to mitigate the barren plateau phenomenon~\cite{larocca2025barren}. Among them, one can find evolutionary-based VQAs training \cite{ACAMPORA2024101756}, a quantum–classical hybrid model that applies evolutionary optimization to train variational quantum circuits, leveraging adaptive search to enhance quantum algorithm performance.

In parallel, Quadratic Unconstrained Binary Optimization (QUBO) formulations have also been investigated as a training strategy in Machine Learning. For instance, Date et al.~\cite{Date2021} reformulated classical models such as linear regression, support vector machines (SVMs), and balanced k-means clustering into QUBO problems solvable via adiabatic quantum computing (AQC), reporting comparable or even superior performance with respect to conventional methods. QUBO Formulations for training artificial neural networks \cite{date2021qubo, 10.3389/frai.2024.1368569}, Ising formulations \cite{Lucas_2014},  as well as coherent Ising machines \cite{pramanik2024ising} have also been proposed as hybrid classical-AQC machine learning models. 

Building on this line of research, our previous work introduced the use of QUBO reformulations and AQC as a novel optimization framework for training variational quantum circuits~\cite{10.1007/978-3-031-90200-0_8, acosta2025qubovqa}. To the best of our knowledge, no other QUBO or AQC-based alternative have been proposed in the literature to optimize VQAs based on annealing or adiabatic techniques, apart from our previous work. Particularly, we refer the reader to reference~\cite{acosta2025qubovqa} for a broader application of the proposed model beyond entanglement detection.

Regarding the entanglement detection and characterisation problem, several machine learning techniques have been recently proposed, including  solutions based on classical neural networks \cite{Urena2024,Asif2023} and kernel methods, both classical \cite{casale2023large} and quantum \cite{Mahdian2025}. These approaches require the full knowledge of the quantum state, obtained by quantum tomography or by a theoretical description. From a quantum computing perspective, it is also interesting to develop methods that can work directly with quantum states, coming from experiments or a previous quantum computation. VQAs are a natural candidate to fulfill this purpose, given its widespread adoption in classification applications~\cite{Cerezo2021}.  By the use of a geometric measure, VQAs can separate entangled from separable states for dimensions up to $25$ qubits \cite{Wang2022,Munoz-Moller2022}. Unfortunately, the proposed techniques require that the training of the VQA is performed for each state to classify, giving unbearable computing times for big systems. 

In this paper, we explore the use of adiabatic-based optimization of VQAs to address the Quantum Separability problem. 

The rest of the paper is structured as follows. Section~\ref{problem} formalizes the separability problem, describing the mathematical representation of quantum states and the inherent difficulty in distinguishing entangled from separable ones. Section~\ref{methods} summarizes the adiabatic training strategy proposed for this problem, briefly describing the proposed search space discretization, hierarchical segmentation, and QUBO-based formulations. Finally, Section~\ref{results} compares the adiabatic method against classical neural networks and gradient descent-based methods for training VQAs.

\section{The separability problem}\label{problem}

The state of a quantum system is defined as a vector of a complex vector space---usually referred to as a \defn{Hilbert} space and denoted as $\cH$--- of dimension $d$. We use the notation    $\ket{\psi}\in\cH$, where $\ket{\cdot}$ is  a column vector. Additionally,  row vectors---denoted by $\bra{\cdot}$---correspond to linear functionals, i.e., $\bra{\cdot}:\cH\to\C$.   Defining an orthonormal basis, $\{\ket{i}\, \vert\, i\in(0,\ldots,d-1\}\in\cH$, any state vector $\ket{\psi}\in\cH$ can be expressed with respect to this basis as
    \begin{equation}
        \ket{\psi} = \sum_{i=0}^{d-1}\, \psi_i\, \ket{i}\, ,
        \label{eq:vecexpansion}
    \end{equation}
where $\psi_i=\braket{i}{\psi}\in\mathbb{C}$, represents the inner product between vectors $\ket{i}$ and $\ket{\psi}$.

Entanglement is a property of composite quantum systems. Consider a system composed of N quantum subsystems, each with its own Hilbert space $\cH_n,\; n\in(1,\ldots,N)$.
The Hilbert space corresponding to the composite system is given by the \defn{tensor product} of the individual subsystem Hilbert spaces:
    \begin{equation}
        \cH = \cH_1\otimes\cH_2\otimes\ldots\otimes\cH_N:=\bigotimes_{n=1}^N \cH_{n}\,.
    \label{eq:tensorproduct}
    \end{equation}

Any state of the composite Hilbert space $\cH$ can be decomposed in the form 
\begin{equation}
    \ket{\psi}= \sum_{i=1}^N \ket{\psi}_1 \otimes \ket{\psi}_2 \otimes \cdots \otimes \ket{\psi}_N,
    \label{eq:statestensorproduct}
\end{equation}
where $\psi_i\in\cH_i$.

The separability problem can be formulated as follow: Given an arbitrary quantum state $\ket{\psi}$, determine if it can be decomposed in the form 

\begin{equation}
    \ket{\psi}= \ket{\psi}_1 \otimes \ket{\psi}_2 \otimes \cdots \otimes \ket{\psi}_N,
    \label{eq:separable}
\end{equation}
for a certain basis. If this is the case, the system is considered \defn{separable}, otherwise it is \defn{entangled}. This is an NP-hard problem in general. For systems of dimension up to 
$2 \times 3$ the Peres-Horodecki criteria provides both sufficient and necessary conditions to determine if the system is entangled or separable \cite{peres:prl96,horodecki:pla96}. On the contrary, for systems with higher dimension there is no general necessary conditions known.

\section{VQA training with Adiabatic Quantum Computers}\label{methods}

The proposed methodology builds on the work of Acosta et al., whose detailed description can be found in~\cite{10.1007/978-3-031-90200-0_8, acosta2025qubovqa}. Here, we briefly outline the methodology and apply it to the entanglement detection problem. 

\subsection{QUBO-Based model}

The distinctive feature of the proposed approach is the use of an adiabatic quantum computer to perform the optimization of the VQA parameters, thereby leveraging quantum annealing techniques via a QUBO formulation.

In a QUBO formulation the problem is expressed using binary variables and a quadratic objective function to be minimized:
\begin{equation}
\sum_i C_{ii} x_i + \sum_{i<j} C_{ij} x_i x_j
\label{eq:general_qubo}
\end{equation}

where:
\begin{itemize}
    \item $x_i$ are the binary decision variables.
    \item $C_{ii}$ are coefficients for single variables (\textit{linear terms}).
    \item $C_{ij}$ are coefficients for interactions between pairs of variables (\textit{quadratic terms}).
\end{itemize}

To obtain a proper QUBO formulation, the proposed model requires discretizing the search space, namely the continuous range of possible rotation angles in the range (\(0\) - \(2\pi\)). The target optimization function is based on the MSE (Mean Squared Error) which calculates the mean difference between the target ($y_i$) and calculated ($\hat{y}_i$) state vectors for each of the $r$ training records in a selected dataset:

\begin{equation}
\text{MSE} = \frac{1}{r} \sum_{i=0}^{r-1} \left( y_i - \hat{y}_i \right)^2
\label{eq:mse}
\end{equation}

Let \(\psi_{ij}\) denote the state vector of the target value and \(\phi_{ij}\) the state vector of the calculated value for each of the \(j\) possible states of a system composed of \(q\) qubits. This yields the following QUBO formulation:

\begin{equation}
\mathrm{QUBO} = \frac{1}{r} \sum_{i=0}^{r - 1} \sum_{j=0}^{2^q - 1} \left( \psi_{ij} - \phi_{ij} \right)^2
\label{eq:qubo_total_loss}
\end{equation}

Here, calculated \(\phi_{ij}\) values are defined in symbolic form as:

\begin{equation}
|\phi\rangle = Q |\psi\rangle = \sum_{j=0}^{2^q - 1} \left( \sum_{k=0}^{2^q - 1} Q_{jk} \psi_k \right) |j\rangle
\label{eq:unitary_action}
\end{equation}

where $Q$ is the quantum operator corresponding to the ansatz circuit, expanded in terms of the \(w\) validation points that lead to each of the \(C_{ii}, x_i, C_{ij}, x_j\) coefficients and binary variables in the general QUBO formulation (Eq. \ref{eq:general_qubo})  and  $|j\rangle$ is the \textit{j-th} orthogonal basis of the quantum system. This results in the final QUBO formulation:
\begin{equation}
\mathrm{QUBO} = \frac{1}{r} \sum_{i=0}^{r - 1} \sum_{j=0}^{2^q - 1} \left( \psi_{ij} - \sum_{k=0}^{2^q - 1} Q_{jk} \psi_{ik} \right)^2
\label{eq:final_qubo}
\end{equation}

to be executed on the Adiabatic Quantum Annealer. This process seeks the global minimum of the cost function by gradually evolving the quantum system toward its lowest-energy configuration, thereby potentially alleviating challenges such as barren plateaus that arise in classical gradient-based methods \cite{larocca2025barren}.

\subsection{Hierarchical segmentation of the search space.}

To reduce the computational cost of an exhaustive search, we propose a hierarchical strategy to sample the space of potential solutions with a balance between exploration and exploitation. This approach falls within the family of hierarchical (or multilevel) metaheuristics~\cite{valejo2021critical, luke2013essentials}, iteratively refines the search space by partitioning it into subregions and directing subsequent searches toward promising areas identified in earlier iterations. By combining an exploration phase, which systematically evaluates discretized points across partitions, with an exploitation phase, which concentrates on subregions around the best solutions found, it achieves a balance between global search and local intensification. This adaptive discretization and incremental refinement strategy is closely related to multilevel optimization methods and shares conceptual similarities with variable neighborhood search techniques~\cite{mladenovic1997variable}.

\begin{figure}
    \centering
    \includegraphics[width=0.7\linewidth]{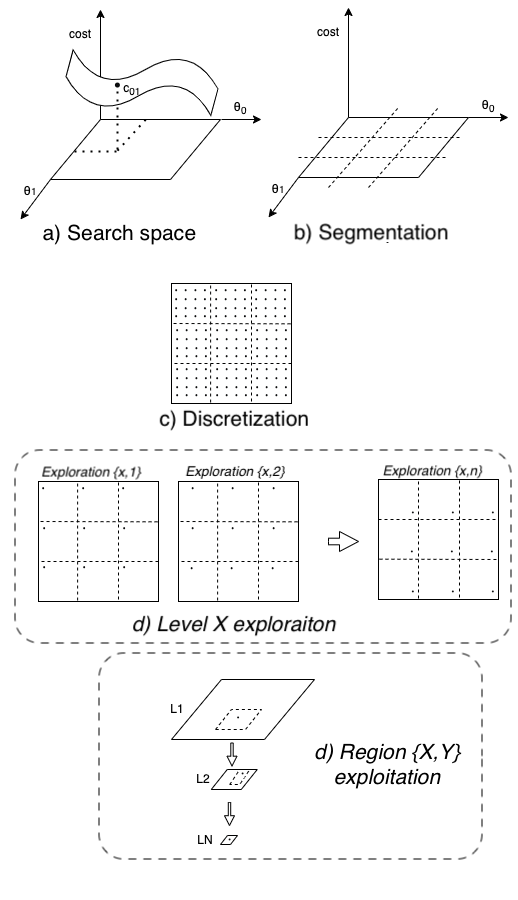}
    \caption{Hierarchical segmentation and discretization process.  a) Search space and corresponding cost function for \(q=2\) parameters, b) Search space segmentation into \(p^q\) regions, c) Discretization map in \(w=3\) points per parameter and segment, d) Level 1 exploration process, e) Level 1 through N exploitation process around region X.}
    \label{fig:iterative-model}
\end{figure}

A schematic representation of the proposed approach is presented in Figure~\ref{fig:iterative-model}, for a 2-parameters model.  Figure ~\ref{fig:iterative-model}.a shows the continuous search space each parameter may take (ie. \(0-2\pi\)) which is divided into \(p\) equally sized partitions generating \(p^q\) \(q-Dimensional\) regions (Fig.~\ref{fig:iterative-model}.b), with \(q\) the number of parameters.  Each region axis is discretized into \(w\) points leading to \(w^q\) discrete evaluation points per region, for instance: \(\{0,pi/12, 2pi/12, 3pi/12, 2pi/3\}\) for first parameter in first region and so on (Fig.~\ref{fig:iterative-model}.c) .  Each of these discrete values have a corresponding binary variable in the QUBO formulation (ie: \(\{x_11, x_12, x_13, ..., x_qq\}\)). The Adiabatic process is executed for each group of evaluation points, that is one point from each region, leading to execution \(E_{lg}\) where \(l\) is the exploration level and \(g\) is the group of evaluation points (Fig.~\ref{fig:iterative-model}.d). Once a group is evaluated, the next group is selected for a new Adiabatic process execution at the same level (\(e_{lg+1}\)) until last group is evaluated. 

Once the Quantum Annealing process is finished at this level, the combination of parameter values leading to the lowest energy configuration in the region is used as pivot point to define the next level's subspace and run the process again in a subspace around this point (Executions \(E_{l+1,g}\)) (Fig.~\ref{fig:iterative-model}.e).

To ensure equal probability across all validation points within the search space, a randomized distribution of level points is performed prior to the onset of the exploration phase. Simultaneously, to accelerate the exploration process, the current level is terminated early if a training step yields an accuracy exceeding either the best result from the previous level or a predefined threshold. This allows the algorithm to promptly proceed to the subsequent level, thereby enhancing overall search efficiency.

Each time a new level is reached, the size of the search space (initially \(0-2\pi\)), and the number of discretization points (\(w\)) can be rescaled by a certain factor.

It is worth mentioning that the number of partitions and discretization points can be adapted to the specific problem and the actual  hardware capacity.

The exploration mechanism, based on simultaneous evaluation of verification points across the entire search space (i.e., one point from each region at the same time) allows, in principle, the adaptive refinement of the search process within a given level by focusing only on regions exhibiting a higher probability of success (e.g., those where the average performance across a set of validation points exceeds a defined threshold, such as 50\%), while discarding regions with consistently low average probability.

The model’s meta-heuristic features (partitioning, discretization, and leveling) enable it to flexibly adapt to varying problem complexities as well as to the capabilities of available hardware. This adaptability allows the model to efficiently run on current NISQ devices using higher partitioning and discretization levels, while also being scalable for future fault-tolerant quantum computers (FTQC) that can operate with lower partitioning and discretization.

This model supports different circuit sizes.  For the sake of clarity, we will hereby describe a simple example for a VQA with 3 qubits and 2 entangling Pauli-Y gates that apply a rotation over the \(Y\) axis on the block sphere by certain angle  which can take values in the range \([0-2\pi)\), leading to a 2-parameter (\(q=2\)) quantum system: \(\theta_0, \theta_1\) (Fig. \ref{fig:iterative-model}.a).  Each angle search space is split in 3 partitions (\(p=3\)): \([0-2\pi/3),[2\pi/3-4\pi/3),[4\pi/3, 2\pi)\) therefore segmenting the original 2D squared search space into 9 regions (Fig. \ref{fig:iterative-model}.b).  Each region's axis \(\theta_x\) is discretized in 4 points (\(w=4\)), leading to a 16 validation points per region (\(w^q\)), 144 validation points in total (Fig. \ref{fig:iterative-model}.c).  At this first level, a QUBO formulation is prepared for a group of validation points including one point per region.  The Adiabatic process will be executed 16 times (\(w^q=4^2\)) validating all the given combinations per execution looking for the lowest energy configuration of the Hamiltonian evolution at each execution (Fig. \ref{fig:iterative-model}.d).  Once the best combination of parameters (validation point) is found, after running all level executions, the search space region is rescaled and a new exploration is started at the next level.  The process is run until the maximum number of levels (\(l=N\)) is reached (Fig. \ref{fig:iterative-model}.e).

\subsection{Scaling}
The parameter search space, discretized into \( v \) points per parameter across \( p \) parameters, typically requires the evaluation of \( v^p \) parameter combinations. The model proposed here leverages adiabatic quantum computation to evaluate a subset \( c \) of parameter combinations simultaneously, thereby reducing the computational complexity to \( v^p / c \). By incorporating a segmentation mechanism of the search space, in which each window consists of \( w \) points per partition and the space is divided into \( s \) partitions such that \( v^p = s \cdot w^p \), the model further reduces execution requirements to \( w^p \) executions, each evaluating \( s^p / c \) combinations. Each execution involving \(p^w\) binary variables, therefore \(p^w\) logical qubits.

When the exploitation phase is executed across \( L \) levels, where each level reduces the search space by a factor \( f_r \) and the number of verification points per partition by a factor \( f_w \), the total number of required executions follows a geometric progression. Specifically, it requires \( w^p \) executions at the first level, \( (w \cdot f_w)^p \) at the second, \( (w \cdot f_w^2)^p \) at the third, and so on. The overall computational cost becomes

\begin{equation}
\sum_{i=0}^{L-1} (w \cdot f_w^i)^p
\label{eq:scaling}
\end{equation}

progressively refining the search with denser sampling at each sublevel.

For instance, a three-dimensional Euclidean search space discretized into 100 points per parameter results in a total of \( 100^3 = 10^6 \) possible combinations. This yields a granularity of \( 0.02\pi \) per step, assuming a full \( 2\pi \) search range. The high dimensionality and resolution of such a space make exhaustive evaluation computationally intensive.  With the proposed model, for 4 levels, and 6 evaluation points per region, and no rescaling factor, the number of combinatorial evaluation points becomes only 864, with the same granularity level of \( 0.02\pi \).  

\section{Experiments and Results}\label{results}

This section presents the methodology for state generation for the quantum entanglement problem and a comparative analysis of performance among different models for tackling this problem: ANNs (Artificial Neural Networks) and VQCs with Classical, Evolutionary and Adiabatic training. 

\subsection{State generation for the quantum entanglement problem}

To study the separability problem we have generated datasets of both separable and entangled states in $2\times 3$ dimensions (three-qubits system). To do so, we use a circuit inspired by the algorithms proposed in Ref. \cite{Urena2024}. As proved in Ref. \cite{Dur00}, a three-qubit state can belong to four different classes regarding its separability: separable, bipartite entangled, $W$-states, and $GHZ$-states. We considered all four classes for the state generation, thereby enabling the trained models to achieve adequate generalization for all possible states for the three-qubit entanglement problem. To create the datasets we use specific random circuits applied to the $\ket{000}$ state in the following way:

\begin{itemize}
\item {\bf Separable states:}  can be prepared just by applying random local unitary transformations to each qubit as displayed in Fig. \ref{fig:circuit_separable}. This state would be given by: 

\begin{equation}
  \begin{array}{c}  
    \ket{\psi}_{\text{sep}} = \left[ U_1(\theta_1,\theta_2,\theta_3) \otimes U_2(\theta_1,\theta_2,\theta_3)  \otimes U_3(\theta_1,\theta_2,\theta_3)  \right]\\ \ket{000}
    \end{array}
\end{equation}
where $U_1(\theta_1,\theta_2,\theta_3)$ is a general one-qubit rotation with Euler angles $(\theta_1,\theta_2,\theta_3)$

\begin{figure}[h!]
    \centering
    \includegraphics[width=0.4\linewidth]{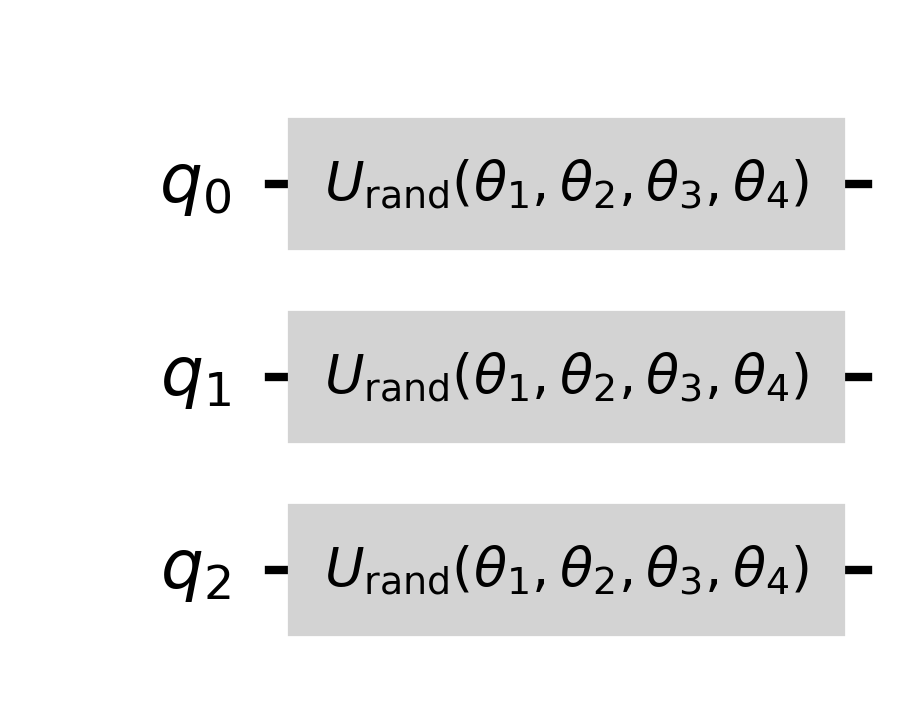}
    \caption{Circuit that generates arbitrary separable states of three qubits. Without loss of generality we select the initial state as $\ket{000}$.}
    \label{fig:circuit_separable}
\end{figure}

\item {\bf Bipartite states: } are designed just by creating a general Bell state in two of the qubits while the other one is an arbitrary single-qubit state, see Fig. \ref{fig:circuit_entangled_bipartite}. 

\begin{equation}
\begin{array}{c}  
    \ket{\psi}_{\text{bip}}= \left[U_1(\theta_1,\theta_2,\theta_3) \otimes U_2(\theta_1,\theta_2,\theta_3)  \otimes U_3(\theta_1,\theta_2,\theta_3)  \right]\\
    (\ket{0}\otimes(\frac{\ket{00}+\ket{11}}{\sqrt{2}}))
\end{array}
\end{equation}

The circuit to construct these states, as seen in Figure \ref{fig:circuit_entangled_bipartite}, starts by creating a Bell pair between two qubits and leaving the third in a separable \(|0\rangle\) state. Local unitary operations \(U_i(\theta_1, \theta_2, \theta_3)\) are applied to all qubits.

\vspace{-10pt}
\begin{figure}[h!]
    \centering
    \includegraphics[width=0.6\linewidth]{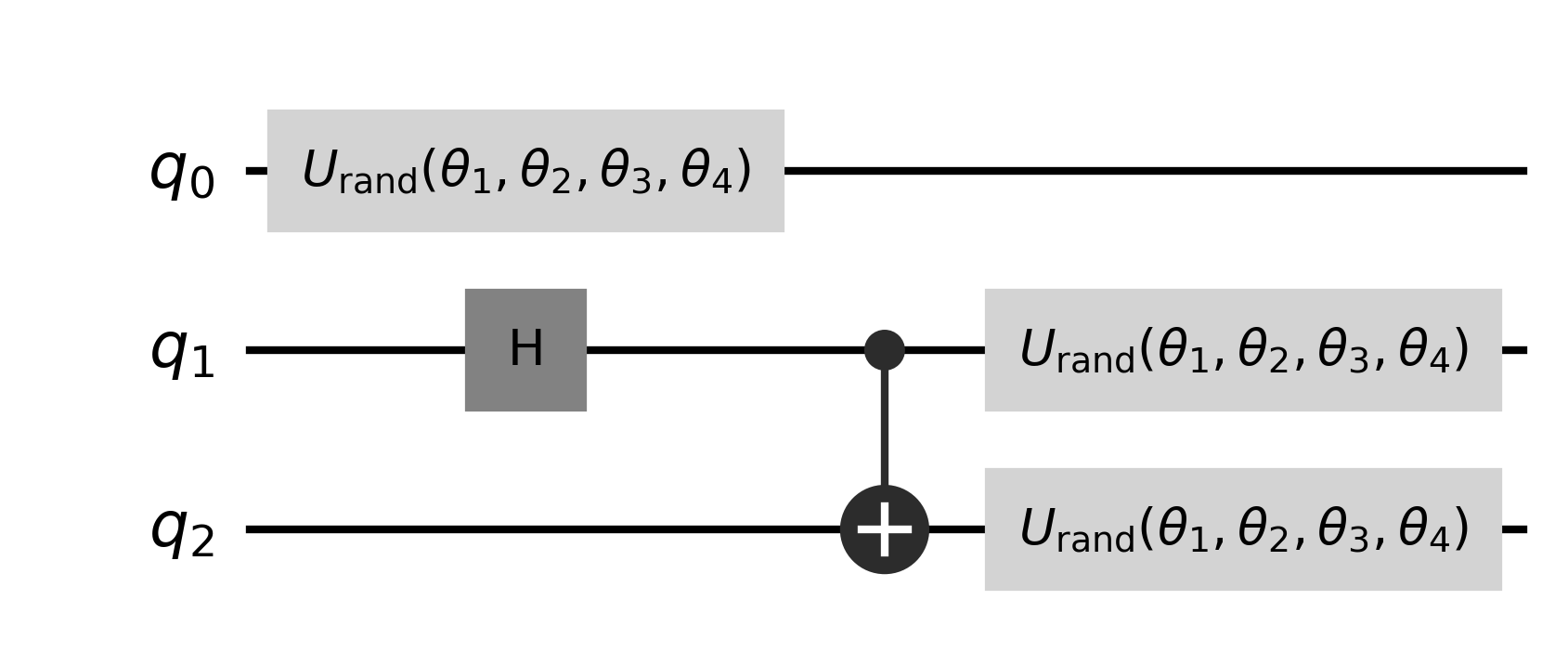}
    \caption{Circuit that generates a Bell pair and an arbitrary single-qubit state.  Without loss of generality we select the initial state as $\ket{000}$.}
    \label{fig:circuit_entangled_bipartite}
\end{figure}
\vspace{-10pt}

\item {\bf W states: } are maximally entangled states with the following form: 
\begin{equation}
\begin{array}{c}  
   \ket{\psi}_W= K_W \left( a\ket{001} + b \ket{010} + c\ket{100} + d\ket{000} \right)
\end{array}
\end{equation}
for arbitrary values of $a,\, b, \, c,\, d$ and being $K_W=1/\sqrt{a^2+b^2+c^2+d^2}$ a normalisation constant. To create this kind of state we can use the circuit seen in Figure \ref{fig:circuit_entangled_w} that encodes the target amplitudes into a four-dimensional subspace of the three-qubit Hilbert space by decomposing the target superposition into unitary operations acting on the initial state \(|000\rangle\).

\vspace{-10pt}
\begin{figure}[h!]
    \centering
    \includegraphics[width=0.45\linewidth]{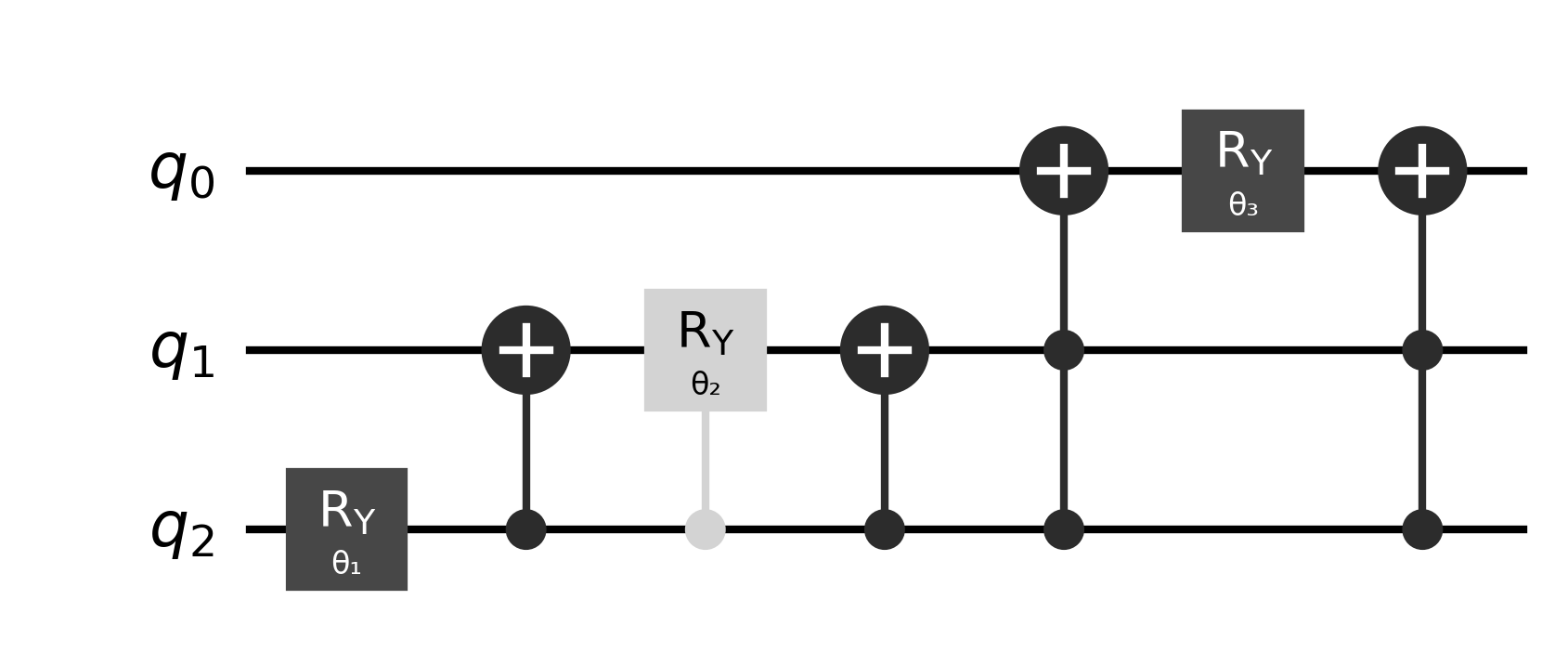}
    \caption{Circuit that generates an entangled W state.  Without loss of generality we select the initial state as $\ket{000}$.}
    \label{fig:circuit_entangled_w}
\end{figure}
\vspace{-10pt}

\item {\bf GHZ states: }  is another family of maximally entangled states, with an expression 
\begin{multline}
\ket{\Psi_{\text{GHZ}}} = \sqrt{K_{GHZ}} \Big( 
\cos(\delta) \ket{0} \ket{0} \ket{0} \\
+ \sin(\delta) e^{i\phi} \ket{\varphi_A} \ket{\varphi_B} \ket{\varphi_C} \Big)
\end{multline}

\begin{align*}
 \ket{\varphi_A} = \cos(\alpha) \ket{0} + \sin(\alpha) \ket{1} \nonumber\\
 \ket{\varphi_B} = \cos(\beta) \ket{0} + \sin(\beta) \ket{1} \nonumber\\
 \ket{\varphi_C} = \cos(\gamma) \ket{0} + \sin(\gamma) \ket{1},
\end{align*}
and $K_{GHZ}=1/(1+ \cos(\delta)\sin(\delta) \cos(\alpha) \cos(\beta)  \cos(\varphi) )$.\\

The circuit, as seen in Figure \ref{fig:circuit_etangled_ghz}, prepares an entangled state by applying first an \( R_y(2\delta) \) gate to the first qubit to encode the amplitudes \( \cos(\delta) \) and \( \sin(\delta) \), followed by a phase gate \( P(\phi) \) to introduce the complex phase. The conditional product state was prepared using controlled \( R_y \) rotations on the remaining qubits, each parameterized by angles \( \alpha \), \( \beta \), and \( \gamma \), such that the rotations only occur when the first qubit is in state \( \ket{1} \). 

\vspace{-10pt}
\begin{figure}[h!]
    \centering
    \includegraphics[width=0.6\linewidth]{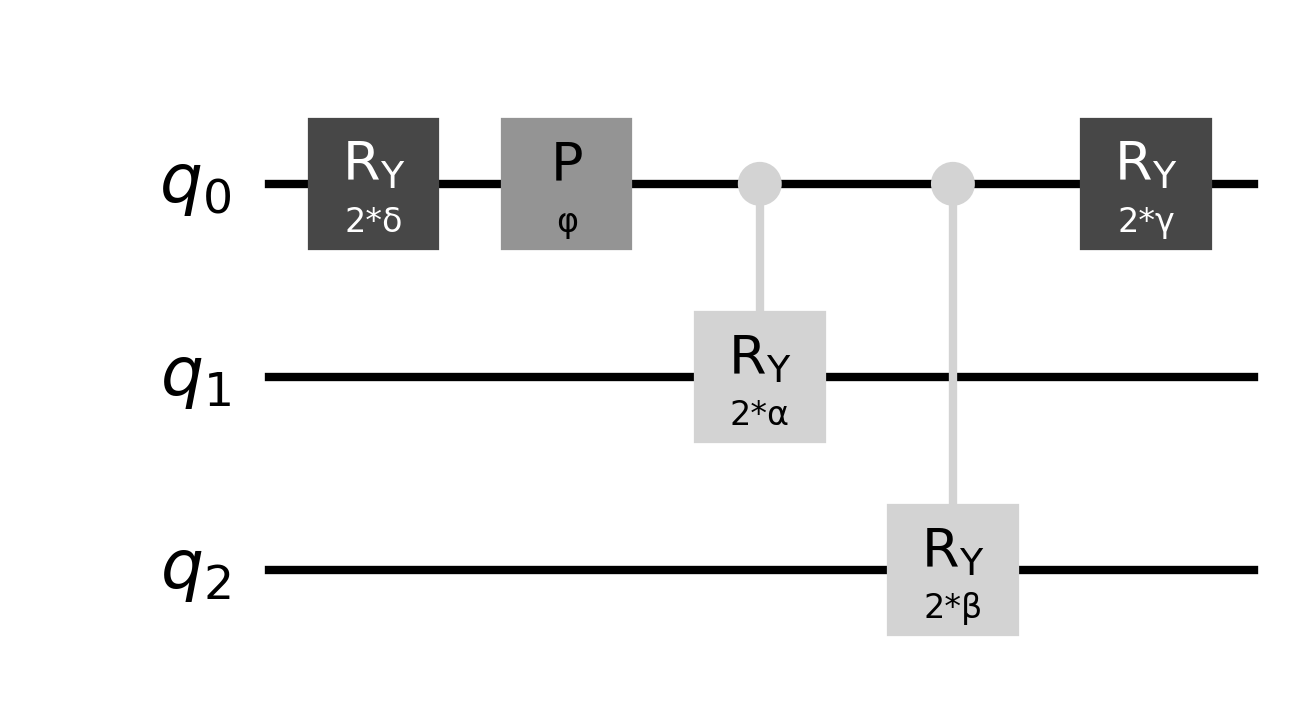}
    \caption{Entangled GHZ, three qubits (\(q_0\) through \(q_2\)).}
    \label{fig:circuit_etangled_ghz}
\end{figure}
\vspace{-10pt}

\end{itemize}

The resulting dataset, shown in Table \ref{tab:dataset}, contains a balanced representation of each class: Separable and Entangled (labeled \(0\) and \(1\), respectively). Within the Entangled class, records corresponding to each of the subclasses (Bipartite, GHZ and W) are equally balanced as well. 

\begin{table}[htbp]
    \caption{Sample Dataset balancing for 96 records  }
    \begin{center}
    \begin{tabular}{|c|c|c|}
    \hline
    \textbf{Class} & \textbf{Label} & \textbf{Num Records} \\
    \hline
    Separable & 0 & 48 \\
    Entangled - Bipartite & 1 & 16 \\
    Entangled - GHZ & 1 & 16 \\
    Entangled - W & 1 & 16 \\

    \hline

    \end{tabular}
    \label{tab:dataset}
    \end{center}
\end{table}

\subsection{Training models settings}

The classical training of the VQCs was performed using SPSA \cite{spall1992spsa} with a number of 200 epochs per training cycle. For the Evolutionary Training model, we tested the Memetic Algorithm \cite{moscato1989memetic} for 10 generations, Tournament Selection and Gaussian Mutation with a sigma factor of 0.2 and a mutation probability of 15\%.  The Artificial Neural Network (ANN) was designed as a feedforward model consisting of an input layer with 3 neurons, a hidden layer of 8 neurons using the ReLU activation function, and an output layer with a single neuron activated by a Sigmoid function, trained using ADAM optimizer.  For the Adiabatic training model, 4 exploitation levels were defined in all cases, with 3 partitions for each angle and 3 validation points per angle. A rescaling factor of 50\% was applied to each level, and 20\% rescaling on the number of data points per angle per level. Each model was run for 25 executions. The code and datasets are publicly available at: \url{https://github.com/eacostam/gradientfreeVQA.git}.

The server configuration used for the experiments was as follows: 32 CPU cores (AMD EPYC 7351 16-Core Processor, 2400 MHz), 33.6 GB of RAM, and the operating system was AlmaLinux 8.8 (Sapphire Caracal, based on RHEL/CentOS/Fedora).

\subsection{Results}

\begin{figure}[h]
    \centering
    \begin{tikzpicture}
    \begin{axis}[
        boxplot/draw direction=y,
        ylabel={Accuracy},
        xtick={1,2,3,4,5,6},
        xticklabels={Ad-t, Ad-v, Cl-t, Cl-v, ANN-t, ANN-v},
        width=0.5\textwidth,
        height=6cm, 
        legend style={at={(0.5,-0.15)}, anchor=north, legend columns=-1}
    ]
    
    \addplot+[
        boxplot,
        fill=blue!20,
        draw=blue
    ] table[y=Value, col sep=comma] {adiab_boxplot_training.csv};

    \addplot+[
        boxplot,
        fill=blue!20,
        draw=blue
    ] table[y=Value, col sep=comma] {adiab_boxplot_validation.csv};
    
    \addplot+[
        boxplot,
        fill=green!20,
        draw=green
    ] table[y=Value, col sep=comma] {classic_boxplot_training.csv};

    \addplot+[
        boxplot,
        fill=green!20,
        draw=green
    ] table[y=Value, col sep=comma] {classic_boxplot_validation.csv};

    \addplot+[
        boxplot,
        fill=gray!20,
        draw=gray
    ] table[y=Value, col sep=comma] {ann_boxplot_training.csv};
    
    \addplot+[
        boxplot,
        fill=gray!20,
        draw=gray
    ] table[y=Value, col sep=comma] 
    {ann_boxplot_validation.csv};
    
    \end{axis}
    \end{tikzpicture}
    \caption{Average accuracy per model (\textbf{Ad}iabatic, \textbf{Cl}assic, \textbf{ANN}) and phase (\textbf{t}raining, \textbf{v}alidation) }.
    \label{fig:box_stats}
\end{figure}
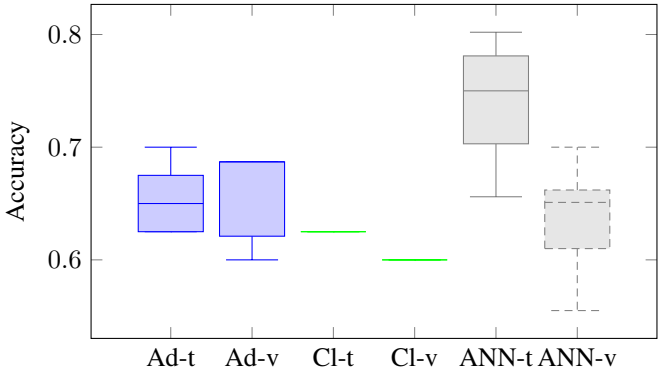

The distributions of results for the different models under consideration can be seen in Figure \ref{fig:box_stats}. 

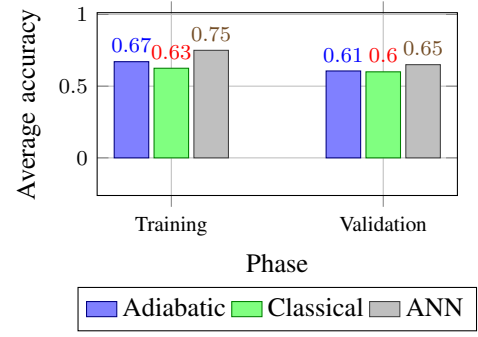
\begin{figure}[h!]
    \centering
    \begin{tikzpicture}
        \begin{axis}[ybar,bar width=13pt,width=0.35\textwidth,
            height=4cm,xlabel={Phase},ylabel={Average accuracy},
            symbolic x coords={Training,Validation},
            xtick=data,ymin=0,grid=major,tick label style={font=\footnotesize},
            enlargelimits=0.35,
            legend style={at={(0.5,-0.5)}, anchor=north, legend columns=3},
            legend image code/.code={\draw[#1,fill=#1] (0cm,-0.1cm) rectangle (0.4cm,0.1cm);}
        ]

        \addplot+[
            fill=blue!50,
            draw=blue!50!black,
            bar shift=-15pt,
            nodes near coords,
            every node near coord/.append style={font=\footnotesize}
        ] table [x=phase,y=accuracy,col sep=comma] {adiabatic_entanglement_results.csv};
        
        \addplot+[
            fill=green!50,
            draw=green!50!black,
            bar shift=0pt,
            nodes near coords,
            every node near coord/.append style={font=\footnotesize}
        ] table [x=phase,y=accuracy,col sep=comma] {classic_entanglement_results.csv};

        \addplot+[
            fill=gray!50,
            draw=gray!50!black,
            bar shift=15pt,
            nodes near coords,
            every node near coord/.append style={font=\footnotesize}
        ] table [x=phase,y=accuracy,col sep=comma] {ann_entanglement_results.csv};

        \legend{Adiabatic, Classical, ANN}
        \end{axis}
    \end{tikzpicture}
    \caption{Compared performance of Adiabatic training of VQAs, Classical training of VQAs and classical ANN models for the entanglement detection problem.}
    \label{fig:classical_ann_training}
\end{figure}

The ANN model demonstrates superior performance compared to quantum-based approaches, achieving consistently stable accuracy levels averaging  75\% on the training dataset and 65\% on the validation dataset, as illustrated in Figure~\ref{fig:classical_ann_training}.

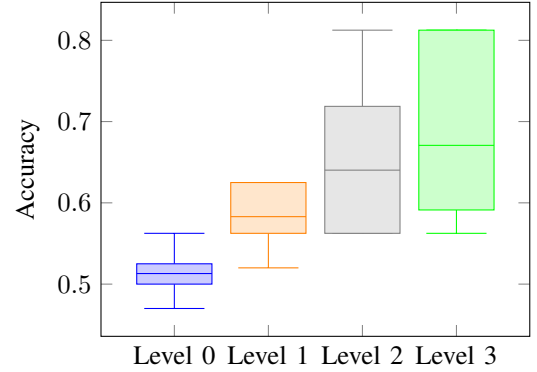
\begin{figure}[h]
    \centering
    \begin{tikzpicture}
    \begin{axis}[
        boxplot/draw direction=y,
        ylabel={Accuracy},
        xtick={1,2,3,4},
        xticklabels={Level 0, Level 1, Level 2, Level 3},
        width=0.4\textwidth,
        height=6cm, 
        legend style={at={(0.5,-0.15)}, anchor=north, legend columns=-1}
    ]
    
    \addplot+[
        boxplot,
        fill=blue!20,
        draw=blue
    ] table[y=Value, col sep=comma] {adiab_boxplot_stats_level0.csv};
    
    \addplot+[
        boxplot,
        fill=orange!20,
        draw=orange
    ] table[y=Value, col sep=comma] {adiab_boxplot_stats_level1.csv};
    
    \addplot+[
        boxplot,
        fill=gray!20,
        draw=gray
    ] table[y=Value, col sep=comma] {adiab_boxplot_stats_level2.csv};

    \addplot+[
        boxplot,
        fill=green!20,
        draw=green
    ] table[y=Value, col sep=comma] {adiab_boxplot_stats_level3.csv};
    
    \end{axis}
    \end{tikzpicture}
    \caption{Training Accuracy by level on Adiabatic training.}
    \label{fig:accuracies}
\end{figure}

The adiabatic quantum model exhibits an average of 67\% on the training phase and 61\% on the validation phase, as shown in Figure~\ref{fig:classical_ann_training}. 
A breakdown by entangled subclass reveals varying degrees of classification difficulty: Bipartite entangled states are identified with higher accuracy (65\%), while GHZ and W states yield lower accuracies (60\% and 55\%, respectively), indicating that W states are the most challenging to distinguish from separable ones in this setting. We also tested that the adiabatic quantum search strategy yields improved performance at each successive hierarchical level, driven by the targeted refinement introduced during the localized exploitation phase as seen on Fig.~\ref{fig:accuracies} for a total of 10 experiments.

The adiabatic optimization approach can achieve comparable levels of accuracy to conventional VQA training methods, as shown in Fig.~\ref{fig:classical_ann_training}. While execution time comparisons remain to be fully evaluated, 
the translation of the VQA training into an adiabatic framework introduces several advantages, including a fixed and potentially reduced number of computational steps independent of data volume, as the adiabatic process explores the entire state space in a single run. 
This process enables a high-level scan of the solution space and benefits from improved quantum hardware, potentially reducing computational costs over time through coarser initial discretization and fewer refinement levels.
On the other hand, the discretization of the Euclidean space introduces a loss of fine-grained resolution and a non-negligible probability of failing to locate the global minimum in highly complex cost functions.







We can observe that VQAs solutions show very competitive performance in comparison with ANNs. Considering that VQAs work with quantum data, this result is very interesting, as these models do not require the full classical knowledge of the state as ANNs-based models do. For practical purposes in quantum computing, this result opens a new possibility of automatic entanglement detection without tomography. 

Due to its substantial computational demands, further work is required to run the baseline evolutionary model for the proposed problem, since this model exceeded the capacity of our current experimental infrastructure. Comprehensive performance metrics will be presented following more extensive testing in a suitable computational environment.

\section{Conclusions}

In this work, we have explored an adiabatic-based optimization framework for training variational quantum algorithms (VQAs) applied to the Quantum Separability problem. Building upon our previous studies on QUBO reformulations and adiabatic quantum computation (AQC) for VQA training, we proposed a novel method capable of directly processing quantum states without requiring full state tomography. This approach extends the applicability of VQAs to large systems by mitigating the computational overhead typically associated with state-by-state training, while providing a complementary alternative to classical and quantum machine learning techniques previously employed for entanglement detection and characterization.

While the discretization of the search space required for the proposed solution inherently introduces a loss of precision, our preliminary results suggest that an appropriate combination of partitioning granularity, evaluation density, and levels of local refinement can effectively approximate the target accuracy. This is accomplished through successive iterations, each requiring significantly reduced computational resources per execution. The proposed model seeks to balance computational efficiency with optimization precision, potentially addressing some of the known limitations of traditional VQA training techniques. By presenting these early findings, we aim to open a dialogue within the scientific community to validate, refine, and ultimately mature the methodology into a robust and scalable solution for adiabatic VQA training.

Future research will involve more extensive experimentation including the evolutionary training approach for the entanglement problem, as well as the incorporation of analytical or stochastic decision-making mechanisms to further optimize the search process at each level. 

\section*{Acknowledgment}
We acknowledge funding from MCUI/AEI (10.13039/501100011033) of the Spanish Government and FEDER, EU, 
through project PID2021-128970OA-I00 and QUANTUM ENIA project call - Quantum Spain project, and by the EU through the Recovery, Transformation and Resilience Plan - NextGenerationEU within the framework of the Digital Spain 2026 Agenda. We also acknowledge project C-EXP-251-UGR23, cofunded by Consejería de Universidad, Investigación e Innovación and European Union call FEDER Andalucía 2021-2027. Besides, this work was supported by grants PID2021-123041OB-I00 funded by MCIN/AEI/ 10.13039/501100011033 and by “ERDF A way of making Europe” and by grant PID2024-158203OB-I00 funded by MICIU/AEI/10.13039/501100011033 and ERDF/EU.

\bibliographystyle{IEEEtran}
\bibliography{QAI_VQA_AQC_entanglement}  

@article{Dur00,
    title = {Three qubits can be entangled in two inequivalent ways},
    author = {D\"ur, W. and Vidal, G. and Cirac, J.~I.},
    journal = {Phys. Rev. A},
    volume = {62},
    issue = {6},
    pages = {062314},
    numpages = {12},
    year = {2000},
    month = {Nov},
    publisher = {American Physical Society},
    doi = {10.1103/PhysRevA.62.062314},
    url = {https://link.aps.org/doi/10.1103/PhysRevA.62.062314}
}

@article{casale2023large,
  title={Large-Scale Quantum Separability Through a Reproducible Machine Learning Lens},
  author={Casal{\'e}, Balthazar and Di Molfetta, Giuseppe and Anthoine, Sandrine and Kadri, Hachem},
  journal={arXiv preprint arXiv:2306.09444},
  year={2023}
}

@article{Cerezo2021,
  author  = {M. Cerezo and Andrew Arrasmith and Ryan Babbush and Simon C. Benjamin and Suguru Endo and Keisuke Fujii and Jarrod R. McClean and Kosuke Mitarai and Xiao Yuan and Lukasz Cincio and Patrick J. Coles},
  title   = {Variational quantum algorithms},
  journal = {Nature Reviews Physics},
  volume  = {3},
  pages   = {625--644},
  year    = {2021},
  month   = {August},
  doi     = {10.1038/s42254-021-00348-9},
  url     = {https://doi.org/10.1038/s42254-021-00348-9}
}

@article{einstein:pr35,
	author = {Einstein, A. and Podolsky, B. and Rosen, N.},
	doi = {10.1103/PhysRev.47.777},
	journal = {Phys. Rev.},
	number = {10},
	numpages = {3},
	pages = {777--780},
	publisher = {American Physical Society},
	title = {Can Quantum-Mechanical Description of Physical Reality Be Considered Complete?},
	volume = {47},
	year = {1935}}

@article{schrodinger:mpcps35,
	author = {E. Schr\"odinger},
	journal = {Math. Proc. Cam. Phil. Soc.},
	pages = {555},
	title = {Discussion of Probability Relations between Separated Systems},
	volume = {31},
	year = {1935},
}

@article{horodecki:rmp09,
	author = {R. Horodecki and P. Horodecki and M. Horodecki and K. Horodecki},
	journal = RMP,
	pages = {865--942},
	title = {Quantum Entanglement},
	volume = {2009},
	year = 2009,
	}

@article{peres:prl96,
	author = {Peres, A.},
	doi = {10.1103/PhysRevLett.77.1413},
	journal = PRL,
	number = {8},
	numpages = {2},
	pages = {1413--1415},
	publisher = {American Physical Society},
	title = {Separability Criterion for Density Matrices},
	volume = {77},
	year = {1996}}

@article{horodecki:pla96,
	author = {M. Horodecki AND P. Horodecki AND R. Horodecki},
	issn = {0375-9601},
	journal = {Physics Letters A},
	number = {1-2},
	pages = {1 - 8},
	title = {Separability of mixed states: necessary and sufficient conditions},
	volume = {223},
	year = {1996},
	}

@article{Urena2024,
  author  = {Julio Ureña and Antonio Sojo and Juani Bermejo-Vega and Daniel Manzano},
  title   = {Entanglement detection with classical deep neural networks},
  journal = {Scientific Reports},
  volume  = {14},
  pages   = {18109},
  year    = {2024},
  doi     = {10.1038/s41598-024-43980-4},
  url     = {https://doi.org/10.1038/s41598-024-43980-4}
}

@article{Asif2023,
  author  = {Naema Asif and Uman Khalid and Awais Khan and Trung Q. Duong and Hyundong Shin},
  title   = {Entanglement detection with artificial neural networks},
  journal = {Scientific Reports},
  volume  = {13},
  pages   = {1562},
  year    = {2023},
  doi     = {10.1038/s41598-023-28629-w},
  url     = {https://doi.org/10.1038/s41598-023-28629-w}
}

@article{Wang2022,
  author  = {Kun Wang and Zhixin Song and Xuanqiang Zhao and Zihe Wang and Xin Wang},
  title   = {Detecting and quantifying entanglement on near-term quantum devices},
  journal = {npj Quantum Information},
  volume  = {8},
  pages   = {52},
  year    = {2022},
  doi     = {10.1038/s41534-022-00568-2},
  url     = {https://doi.org/10.1038/s41534-022-00568-2}
}

@article{Munoz-Moller2022,
  author  = {A.D. Muñoz-Moller and L. Pereira and L. Zambrano and J. Cortés-Vega and A. Delgado},
  title   = {Variational Determination of Multiqubit Geometrical Entanglement in Noisy Intermediate-Scale Quantum Computers},
  journal = {Physical Review Applied},
  volume  = {18},
  pages   = {024048},
  year    = {2022},
  month   = {August},
  doi     = {10.1103/PhysRevApplied.18.024048},
  url     = {https://doi.org/10.1103/PhysRevApplied.18.024048}
}

@article{Mahdian2025,
  author  = {Mahmoud Mahdian and Zahra Mousavi},
  title   = {Entanglement detection with quantum support vector machine (QSVM) on near-term quantum devices},
  journal = {Scientific Reports},
  volume  = {15},
  pages   = {11931},
  year    = {2025},
  doi     = {10.1038/s41598-025-09793-9},
  url     = {https://doi.org/10.1038/s41598-025-09793-9}
}

@InProceedings{10.1007/978-3-031-90200-0_8,
author="Acosta, Ernesto
and Guti{\'e}rrez, Carlos Cano
and Botella, Guillermo
and Campos, Roberto",
editor="Caino-Lores, Silvina
and Zeinalipour, Demetris
and Doudali, Thaleia Dimitra
and Singh, David E.
and Garz{\'o}n, Gracia Ester Mart{\'i}n
and Sousa, Leonel
and Andrade, Diego
and Cucinotta, Tommaso
and D'Ambrosio, Donato
and Diehl, Patrick
and Dolz, Manuel F.
and Jukan, Admela
and Montella, Raffaele
and Nardelli, Matteo
and Garcia-Gasulla, Marta
and Neuwirth, Sarah",
title="Adiabatic Training for Variational Quantum Algorithms",
booktitle="Euro-Par 2024: Parallel Processing Workshops",
year="2025",
publisher="Springer Nature Switzerland",
address="Cham",
pages="88--99",
isbn="978-3-031-90200-0"
}

@article{ACAMPORA2024101756,
title = {EVOVAQ: EVOlutionary algorithms-based toolbox for VAriational Quantum circuits},
journal = {SoftwareX},
volume = {26},
pages = {101756},
year = {2024},
issn = {2352-7110},
doi = {https://doi.org/10.1016/j.softx.2024.101756},
url = {https://www.sciencedirect.com/science/article/pii/S2352711024001274},
author = {Giovanni Acampora and Carlos {Cano Gutiérrez} and Angela Chiatto and José Manuel {Soto Hidalgo} and Autilia Vitiello}
}

@article{date2021qubo,
  title     = {QUBO formulations for training machine learning models},
  author    = {P. Date, D. Arthur and L. Pusey-Nazzaro},
  journal   = {Scientific Reports},
  volume    = {11},
  pages     = {10029},
  year      = {2021},
  publisher = {Nature Publishing Group},
  doi       = {10.1038/s41598-021-89461-4},
  url       = {https://doi.org/10.1038/s41598-021-89461-4}
}

@inproceedings{pramanik2024ising,
  title     = {Training Machine Learning Models with Ising Machines},
  author    = {Sayantan Pramanik and Kaumudibikash Goswami and Sourav Chatterjee and M. Girish Chandra},
  booktitle = {Proceedings of the ML with New Compute Paradigms Workshop at NeurIPS 2024},
  year      = {2024},
  url       = {https://openreview.net/forum?id=Du5geIL4rM}
}

@article{valejo2021critical,
  title={A Critical Survey of the Multilevel Method in Complex Networks},
  author={Valejo, Andre and Ferreira, Valmir C and Fabbri, Ricardo and de Oliveira, M{\'a}rio C F and de Andrade Lopes, Andr{\'e}},
  journal={ACM Computing Surveys},
  volume={54},
  number={2},
  pages={1--36},
  year={2021},
  publisher={ACM},
  doi={10.1145/3388460},
  url={https://dl.acm.org/doi/10.1145/3388460}
}

@article{mladenovic1997variable,
  title={Variable Neighborhood Search},
  author={Mladenovi{\'c}, Nenad and Hansen, Pierre},
  journal={Computers \& Operations Research},
  volume={24},
  number={11},
  pages={1097--1100},
  year={1997},
  publisher={Elsevier},
  doi={10.1016/S0305-0548(97)00031-2},
  url={https://doi.org/10.1016/S0305-0548(97)00031-2}
}

@article{Date2021,
  author    = {Prasanna Date and Davis Arthur and Lauren Pusey-Nazzaro},
  title     = {{QUBO} Formulations for Training Machine Learning Models},
  journal   = {Scientific Reports},
  year      = {2021},
  volume    = {11},
  number    = {1},
  pages     = {10029},
  doi       = {10.1038/s41598-021-89461-4},
  url       = {https://doi.org/10.1038/s41598-021-89461-4}
}

@article{10.3389/frai.2024.1368569,
AUTHOR={Abel, Steve  and Criado, Juan Carlos  and Spannowsky, Michael },
TITLE={Training neural networks with universal adiabatic quantum computing},
JOURNAL={Frontiers in Artificial Intelligence},
VOLUME={7},
YEAR={2024},
URL={https://www.frontiersin.org/journals/artificial-intelligence/articles/10.3389/frai.2024.1368569},
DOI={10.3389/frai.2024.1368569},
ISSN={2624-8212}
}

@book{luke2013essentials,
  title     = {Essentials of Metaheuristics},
  author    = {Sean Luke},
  edition   = {Second},
  year      = {2013},
  publisher = {Lulu},
  isbn      = {9781300549628},
  note      = {Available online at \url{http://cs.gmu.edu/~sean/book/metaheuristics/}}
}

@article{larocca2025barren,
  title        = {Barren Plateaus in Variational Quantum Computing},
  author       = {Martín Larocca and Supanut Thanasilp and Samson Wang and Kunal Sharma and Jacob Biamonte and Patrick J. Coles and Lukasz Cincio and Jarrod R. McClean and Zoë Holmes and M. Cerezo},
  journal      = {Nature Reviews Physics},
  volume       = {7},
  pages        = {174--189},
  year         = {2025},
  doi          = {10.1038/s42254-025-00813-9},
  url          = {https://doi.org/10.1038/s42254-025-00813-9}
}

@article{spall1992spsa,
  author  = {Spall, James C.},
  title   = {Multivariate stochastic approximation using a simultaneous perturbation gradient approximation},
  journal = {IEEE Transactions on Automatic Control},
  volume  = {37},
  number  = {3},
  pages   = {332-341},
  year    = {1992},
  doi     = {10.1109/9.119632}
}

@techreport{moscato1989memetic,
  author      = {Pablo Moscato},
  title       = {On Evolution, Search, Optimization, Genetic Algorithms and Martial Arts: Towards Memetic Algorithms},
  year        = {1989},
  institution = {California Institute of Technology},
  number      = {826},
  type        = {Technical Report}
}

@article{Lucas_2014,
   title={Ising formulations of many NP problems},
   volume={2},
   ISSN={2296-424X},
   url={http://dx.doi.org/10.3389/fphy.2014.00005},
   DOI={10.3389/fphy.2014.00005},
   journal={Frontiers in Physics},
   publisher={Frontiers Media SA},
   author={Lucas, Andrew},
   year={2014} }

@misc{acosta2025qubovqa,
      title={QUBO-based training for VQAs on Quantum Annealers}, 
      author={Ernesto Acosta, Guillermo Botella, Carlos Cano},
      year={2025},
      eprint={2509.01821},
      archivePrefix={arXiv},
      primaryClass={quant-ph},
      url={https://arxiv.org/abs/2509.01821}, 
}

\vspace{12pt}

\end{document}